\documentclass[journal]{IEEEtran}
\usepackage{amsmath}
\usepackage{graphicx}
\usepackage{float}
\usepackage{zed-csp}
\usepackage{makecell}
\usepackage{caption}
\usepackage[labelfont=bf]{caption}
\usepackage{subcaption}
\usepackage{adjustbox}
\usepackage{array}
\usepackage{multicol}
\usepackage{multirow}

\begin{document}

\title{A Formal Specification Framework for Smart Grid Components}

\author{Waseem~Akram, and~Muaz~Niazi,~\IEEEmembership{Senior Member, IEEE}\thanks{$^*$Corresponding author muaz.niazi@ieee.org}}


\maketitle

\begin{abstract}
Smart grid can be considered as the next step in the evolution of power systems. It comprises of different entities and objects ranging from  smart appliances, smart meters, generators, smart storages, and more. One key problem in modeling smart grid is that while currently there is a considerable focus on the practical aspects of smart grid, there are very few modeling attempts and even lesser attempts at formalization. To the best of our knowledge, among other formal methods, formal specification has previously not been applied in the domain of smart grid. In this paper, we attempt to bridge this gap by presenting a novel approach to modeling smart grid components using a formal specification approach. We use a state-based formal specification language namely Z (pronounced as `Zed') since we believe Z is particularly suited for modeling smart grid components.We demonstrate the application of Z on key smart grid components. The presented formal specification can be considered as first steps towards modeling of smart grid using a Software Engineering formalism. It also demonstrates how complex systems, such as the smart grid, can be modeled elegantly using formal specification.
 
\end{abstract}

\begin{IEEEkeywords}
Formal specification, Smart grid, Complex Adaptive System

\end{IEEEkeywords}


\section{Introduction}
\label{sec:intro}

A smart grid can be considered as an advanced and radically evolved version of traditional power systems. The term 'smart' in the smart grid exemplifies the use of bi-directional communication, artificial intelligence, Complex systems theory, modeling and simulation, and more, all employed with the goal of converting the legacy power grid into an advanced proactive and reactive system. 
At the lowest level however, the Smart grid can be considered as an integrated system made up of a variety of interacting components - ranging from smart appliances and smart storages to smart generators, Internet of Things (IoT), and beyond. Another key focus of the Smart grid is in the integration of renewable energy resources, such as, but not limited to, wind turbines and solar panels \cite{wong2017opportunities}. 
By integrating advanced communication and information systems, Smart grid components can communicate and coordinate with each other with the goal of constructing a sustainable and efficient energy production system \cite{gungor2011smart} for the future.
 
As the size of smart grid increases, however, it is coupled with an increase in its intrinsic complexity  \cite{cintuglu2017survey}. Smart grid implies complexity\cite{monti2010power}. To understand this complexity, it is necessary to ponder on the fact that, in any modern large-scale power system, each component can itself be dynamic in its very nature. As such, the states of the system, which are a function of and the result of emergent properties of the numerous interacting components, can also vary temporally. The result of what we see at the macro-level cannot thus be easily discernible as a direct function of the micro-level. This behavior can thus be considered as the outcome of numerous interactive events occurring in relation to each component - quite similar to a natural Complex Adaptive Systems (CAS) \cite{amin2005toward}. 
With so much complexity at hand, it is clear that modeling the smart grid certainly needs considerable number of practical examples and case studies \cite{pagani2014power}. It is also evident that developing various types of formalisms for the domain is needed. This will allow for the selection of better and more elegant solution to models - which in turn can be used to develop a better understanding of the complex domain. 
Essentially, modeling any system can be considered as an activity which allows for a better understanding of the system. In the smart grid domain, better modeling approaches can not only simplify system complexity but also allow for a better understanding and implementation of the system. Besides, it can also allow for ensuring a reduction in system failures.

Formal methods provide facilities for the modeling of each component of any complex system \cite{hall1990seven}. It allows for developing models for each component of the system allowing for a clear focus on understanding consistency as well as semantic correctness. The behavior of each system can be analyzed and observed with the help of these formal models. A  key benefit to this approach is that it helps in the detection of faults and flaws in the design phase of system development, thereby considerably improving system reliability.   

In previous studies, formal specification framework has been successfully applied for the mathematical modeling of different CAS ranging across various domains.  Some key examples of such work includes a formal specification used for the modeling of AIDS spread using agent-based modeling  \cite{siddiqa2013novel}. Likewise, it has been developed for modeling the progression of researchers in their domain \cite{hussain2014toward}, and for the modeling of wireless sensor networks \cite{niazi2011novel}. The use of formal specification models for modeling CAS also include studies such as \cite{Zafar2017,afzaal2016formal}. 
Suggestions to use formal specification for the Smart grid have also previously been mentioned in literature \cite{hackenberg2012applying}. Another example is the use of state machine formalism \cite{turner2014formalized}. However, to the best of our knowledge, the same approach has not been applied much in the domain of the smart grid domain. It is thus clear that there is a growing need to model the key components in a smart grid by means of an elegant formal framework among other tools such as noted previously \cite{rohjans2014requirements}. Such a prudent approach allows for a better understanding of the domain besides allowing for systems to be verified using the given specification. 
 
In this paper, we present first steps towards a basic formal specification modeling framework for smart grid components. We first consider different types of entities and then elaborate their detailed formal specifications. 

The rest of the paper is structured as follows: Section 2 provides basic concept of a formal framework and a smart grid scenario is discussed. Section 3 presents formal specification of different entities of smart grid system. The paper ends with a conclusion in section 4.
\section{Theoretical Foundation}
In this section, we present the theoretical foundations needed for understand the presented formal specification model.
\subsection{Formal Frameworks}
A formal specification is a software engineering approach that is used for mathematical modeling of different components of a system \cite{hall1990seven}. During engineering of any system, it is important to ensure that all the system components are integrated and working correctly without any error. A formal specification provides this facility to accurately specify each requirement of the system before going to the real implementation \cite{woodcock1996using}. 

A formal specification decomposes the large system into the subsystem. Then provides a specification for each individual subsystems. It follows two approaches; one is the algebraic approach in which each operation and relationships can be described, second is a model-based approach which concerns with the state and transitions of each individual components. The model-based approach uses the Z language specification which involves the utilization of mathematical notations, sets, sequences, and states \cite{bowen1996formal}.

\subsection{Scenario of the Smart grid}
We can think of a smart grid as a complex system in which different consumers and generation units are connected through power communication lines \cite{milanovic2017modelling}. Generation units generate power and transmit toward consumer's side. Consumers demand energy according to their usage profile and generation units response according to the consumer's demand \cite{yang2017distributed}.

On the consumer's side, the process of monitoring and controlling consumer's demand energy usage profile is called home energy management (HEM) \cite{hosseini2017non}. This process involves the use of smart meters, smart appliances, information and communication technology etc. the smart meters collect data about consumer's demand at the different time period and transmit this collected data to the power generation unit. On the power generation unit side, this data is analyzed and response according to the consumer's demand. Sometimes, consumer's demands exceed from available energy power at the grid side. This creates unbalance situation of the power system. To handle this issue, the concept of demand response management (DRM)\cite{soares2017stochastic} has been introduced. 

DRM is the process of adjusting consumer's demand in response to the price of energy and incentives from grid side. The generation unit sends their energy cost pattern of different time periods. The time in which power cost is high is called high peak hour and the time in which power cost is low is called off peak hour. So during high peak hour, consumers can keep their flexible appliances off and run at low peak hour.
 
To reduce burden on generation unit, the concept of renewable energy sources (RES) is also introduced in a smart grid application \cite{park2017residential}. RES can be in the form of photo-voltaic (PV) or wind energy sources. These energy sources can be installed and provide energy to the consumer's side. Consumers can use this energy at an off-peak hour, so this solves the energy unbalance as well as high-cost issues. The energy collected from RES can be stored using storage devices. Sometimes, if the available energy is larger than consumer's demands, then it can be sold back to the grid unit by using some buyback mechanism \cite{chiu2017optimized}. However, RES has unpredictable nature and depends on weather and time. PV energy can only be produced during the daytime i.e. at sunny day. Wind energy can only be produced in windy environments. 

Regarding RES, a new concept has also been included in a smart grid called electric vehicles (EVs) \cite{ahmadian2017plug}. These EVs are using energy storage devices which can be charged either using PV or electric station. The stored energy is then used for drive vehicles. In case if the energy demands of consumers exceed at grid unit. These EVs can also provide energy to the grid unit by using the vehicle to grid (V2G) mechanism. 

\section{Formal specification Framework}
In this section, we present formal specification models for the smart grid components. In our study, we consider four different entities in a smart grid system. These entities are: appliance, solar, turbine, and storage devices. Each entity has different states as well as associated events that cause state transitions. In table \ref{tbl:componetns}, summary of each object, states and events are given.

\subsection{Smart appliance}
\begin{figure}[H]
\begin{center}
  \includegraphics[width=8.0 cm, height=10.0 cm]{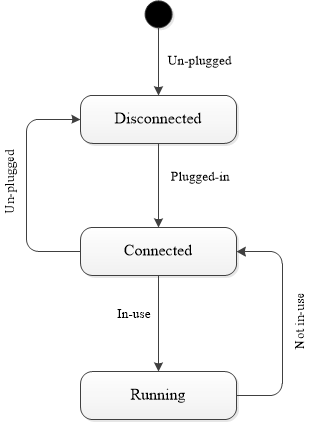}
  \caption{An Appliance lifetime}\label{fig:applifetime}
\end{center}
\end{figure}

An appliance object has three distinct states along with events associated with states. As we can see in figure \ref{fig:applifetime}, Whenever an unplugged event occurs, the appliance enters in the disconnected state. By plugged-in event, the appliance enters in the connected state. And whenever an in-use event occurs, the appliance remains in the running state. 

\begin{table*}[h]
\caption{Smart grid components, their states and events}\label{tbl:componetns}
\centering
\begin{tabular}{| p{2.5cm} | p{4.5cm} | p{4.0cm} |}
\hline
Object & States &	Events  \\\hline

\multirow{3}{*}{Appliance} 
& Disconnected
&Unplugged\\
&Connected &
Plugged-in, not in-use \\
& Running &
In use\\
         
\hline
\multirow{3}{*}{Turbine} 
    & Not running&
No wind\\

&Slow running&
Slow wind\\

&Fast running
&
Fast wind\\
\hline
\multirow{3}{*}{Solar} 
&No energy generation&
Night time\\

&Partial energy generation&
Day, cloudy\\

&Full energy generation&
Day, sunny \\     
         
\hline

\multirow{3}{*}{Storage} 

&Charging&
Store energy\\
&Discharging&
Consume energy\\
&Not-in-use & remove storage\\
\hline
\end{tabular}
\end{table*}

Next, we present formal specification for an appliance object.

A free type ``APPLIANCESTATE'' is used to represent different states of an appliance that are disconnected, connected and running.

\begin{zed}

[APPLIANCESTATE]==\{disconnected,connected,running\}
\end{zed}
Next, we define an \emph{appliance}  schema that contains ``appState'' variable of type ``APPLIANCESTATE''. The value of ``appState'' can be either disconnected, connected or running.
\begin{schema}{Appliance}
appState:APPLIANCESTATE

\end{schema}
As we have defined appliance state and appliance schema. Now we can move towards operational schemas.

First, we start by presenting initialization schema named as \emph{InitAppliance}. In this schema, we declare ``Appliance'' schema as a variable. In predicate section, we say that on initialization, the state of an appliance must be equal to disconnected. 

\begin{schema}{InitAppliance}
Appliance\\
\where
appState=disconnected
\end{schema}

As the plugged-in event occurs, the state of an appliance becomes connected. To show this transition, we define an operational schema called \emph{PluggedInAppliance}. In the schema, the changing state of an appliance is shown by a delta sign$(\Delta)$. In predicate, we say that the state of an appliance is changed from disconnected to the connected state.
\begin{schema}{PluggedInAppliance}
\Delta Appliance\\
\where
appState=disconnected\\
appState'=connected
\end{schema}
The state of an appliance is also changing whenever an in-use event occurs. Then the appliance remains in a running state. This transition is shown by defining \emph{InUseAppliance}. Here, again we declare the changing state of an appliance with a delta sign. In predicate, first an appliance must be in the connected state, then we say that the state changes to the running state.

\begin{schema}{InUseAppliance}
\Delta Appliance
\where
appState=connected\\
appState'=running
\end{schema}
As we noted before, when an unplugged event occurs, an appliance goes to the disconnected state. This transition is shown by another schema called \emph{UnPluggedAppliance}. In the predicate section, we can see that the state of an appliance changes from connected to the disconnected state.

\begin{schema}{UnPluggedAppliance}
\Delta Appliance
\where
appState= connected\\
appState'=disconnected

\end{schema}
An appliance can be in connected but not in use state. This operation is shown by introducing another schema called \emph{NotInUseAppliance}.
\begin{schema}{NotInUseAppliance}
\Delta Appliance
\where
appState= running\\
appState'=connected

\end{schema}

There can be some scenario in which a schema can be success or there may some error. To understanding the success and error conditions of schemas, we present free type definitions in the form of table \ref{tbl:ftypeapp}. In the given table, for each schema we define the success and failure conditions.

\begin{table*}[h]
\centering
\caption{ Success and error outcomes of the schemas for an appliance entity }
\label{tbl:ftypeapp}
\begin{tabular}{| c | c | c |}\hline
\textbf{Schema} & \textbf{Pre-condition for success} &\textbf{Condition for error}\\ \hline  

PluggedInAppliance & \makecell{Appliance is disconnected\\$appState=disconnected$} & \makecell{Already connected\\ $appState=connected$}  \\

\hline

InUseAppliance & \makecell{Appliance is connected \\ 
 $appState=connected$} &\makecell{Already in running state \\
$appState=running$}
\\ \hline 

UnPluggedAppliance & \makecell{Appliance is connected
\\ 
$appState=connected$
\\}  & \makecell{Already in disconnected state \\
$appState=disconnected$
}  \\

\hline
 
NotInUseAppliance & \makecell{Appliance is running\\ $appState = running$\\}  & \makecell{Already not in not running state \\ $appState\neq running$
}  \\

\hline

\end{tabular}

\end{table*}  
\subsection{Wind turbine system}
\begin{figure}[h]
\begin{center}
  \includegraphics[width=8.0 cm, height=10.0 cm]{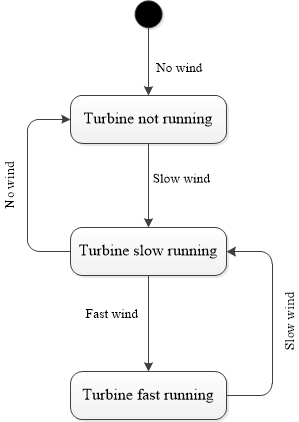}
  \caption{Wind turbine lifetime}\label{fig:turbinelifetime}
\end{center}
\end{figure}

A wind turbine object has three distinct states named as turbine not running, turbine slow running, and turbine fast running. There are also different events associated with each state that causes a state transition. In figure \ref{fig:turbinelifetime} we can see that when there is no wind, the turbine enters in the ``Turbine not running'' state. When there is slow wind, turbine runs slowly. And if there is fast wind, the turbine is running fast.

Next, we present formal specification for a turbine object.

First, we start by defining a free type ``TURBINESTATE'' set. This set comprises of a turbine different states that are ``turbineNotRunning, turbineSlowRunning, and turbineFastRunning''.
\begin{zed}
[TURBINESTATE]==\{trubineNotRunning,\\turbineSlowRunning,turbineFastRunning\}

\end{zed}

Now, we define a wind turbine system schema by presenting \emph{WindTurbine}. This schema consists of a variable ``trbState'' of type \emph{TURBINESTATE}. The value of this variable can be either ``turbineNotRunning'', ``turbineSlowRunning'', or ``turbineFastRunning''.
\begin{schema}{WindTurbine}
trbState:TURBINESTATE
\end{schema}
As we have defined turbine's states and turbine's schema, now we move to operational schemas.

First, we start by initializing a turbine entity by presenting \emph{InitTurbine} schema. In the schema, first, we call the \emph{WindTurbine} schema. Then in the predicate, we say that the initial state of a turbine must be equal to ``turbineNotRunning'' state. 

\begin{schema}{InitTurbine}
WindTurbine
\where
trbState=turbineNotRunning

\end{schema}
As we noted in the state diagram when during slow wind, the state of a turbine becomes ``turbineSlowRunning''. This transition is shown by defining \emph{SlowWind} schema. In the schema, we call the \emph{WindTurbine} schema with a delta sign. In predicate, we say that the state of a turbine is changing from ``turbineNotRunning'' to the ``turbineSlowRunning''.
\begin{schema}{SlowWind}
\Delta WindTurbine
\where
trbState=turbineNotRunning\\
trbState'=turbineSlowRunning

\end{schema}
In case of fast wind, a turbine runs fast. To show this process, we present another schema called \emph{FastWind}. In the schema, again we call \emph{WindTurbine} schema with a delta sign. In predicate, we say that the state of a turbine changes from slow to fast running.
\begin{schema}{FastWind}
\Delta WindTurbine
\where
trbState=turbineSlowRunning\\
trbState'=turbineFastRunning

\end{schema}

A turbine's state also changes to not running when there is no wind. This transition is shown by introducing \emph{NoWind} schema. The schema shows that a turbine's state is changed from slow running to not running state.
\begin{schema}{NoWind}
\Delta WindTurbine
\where
trbState=turbineSlowRunning\\
trbState'=turbineNotRunning

\end{schema}

The schemas of the turbine entity can be success or fail when errors occur. So to understand the success and error conditions of each turbine's schemas, we present free type definitions in table \ref{tbl:ftypetrb}.

\begin{table*}[h]
\centering
\caption{ Success and error outcomes of the schemas for a wind turbine entity }
\label{tbl:ftypetrb}
\begin{tabular}{| c | c | c |}\hline
\textbf{Schema} & \textbf{Pre-condition for success} &\textbf{Condition for error}\\ \hline  

SlowWind & \makecell{Turbine not running\\$trbState=notRunning$} & \makecell{Already in slow running state\\ $trbState=turbineSlowRunning$}  \\

\hline

FastWind & \makecell{Turbine is running slowly \\ 
 $trbState=turbineSlowRunning$} &\makecell{Already in Fast running state \\
$trbState=turbineFastRunning$}
\\ \hline 

NoWind & \makecell{Turbine is running slowly
\\ 
$trbState=turbineSlowRunning$
\\}  & \makecell{Already in not running state \\
$trbState=turbineNotRunning$
}  \\

\hline
 
\end{tabular}
\end{table*}

\subsection{Solar system}
\begin{figure}[h]
\begin{center}
  \includegraphics[width=8.0 cm, height=10.0 cm]{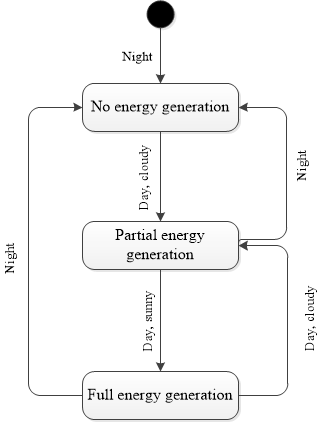}
  \caption{A solar lifetime}\label{fig:solarlifetime}
\end{center}
\end{figure}

A solar object has three distinct states named as ``no energy generation, partial energy generation, and full energy generation''. There are also different events associated with each state that causes a state transition. The states and it's associated events have shown in the figure \ref{fig:solarlifetime}.
Next, we present formal specification for a solar object.

First, we begin by introducing a free type ``SOLARSTATE'' that is used for presenting different states of a solar entity. This set comprises of ``noEnergyGeneration, partialEnergyGeneration, and fullEnergyGeneration'' states.
\begin{zed}
[SOLARSTATE]==\{noEnergyGeneration,\\partialEnergyGeneration  ,fullEnergyGeneration\}
\end{zed}

Here, we present a solar system by introducing \emph{SolarPanel} schema. The schema takes a variable ``slrState'' of type \emph{SOLARSTATE}. The value of this variable can be either ``noEnergyGenreation'', ``partialEnergyGeneration'' or ``fullEnergyGeneration''.

\begin{schema}{SolarPanel}
slrState:SOLARSTATE

\end{schema}

As we have defined solar state and schema, now we move towards the operational schemas.

The initialization process is shown by presenting \emph{InitSolar} schema. First, the schema calls the \emph{SolarPanel} schema in the declaration part. In predicate, we defined that the state of a solar is ``noEnergyGeneration'' at the initial time.
\begin{schema}{InitSolar}
SolarPanel
\where
slrState=noEnergyGeneration
\end{schema}

As we know that when there is a cloud at daytime, then the solar generates partial energy. During this time period, the solar remains in a partial energy generation state. This process is shown by introducing \emph{DayAndCloudy} schema. In the schema, the changing state of a solar is shown by a delta sign. In predicate, it shows that the state of solar changes from no generation to the partial generation.
\begin{schema}{DayAndCloudy}
\Delta SolarPanel
\where
slrState=noEnergyGeneration\\
slrState'=partialEnergyGeneration

\end{schema}
In case of the sunny day, a solar remains in a full generation state. This process is shown by defining \emph{DayAndSunny} schema. Here, we again define the changing state of a solar with delta sign. In predicate, we say that the state of solar changes from partial to full energy generation. 
\begin{schema}{DayAndSunny}
\Delta SolarPanel
\where
slrState=partialEnergyGeneration\\
slrState'=fullEnergyGeneration

\end{schema}

As we know that a solar does not generate any energy during night time. At this time period, solar remains in a no generation state. Here, we present another schema called ``Night''. In the schema, we say that the current state of a solar can be partial or full energy generation. Then we change solar state to the no energy generation state.

\begin{schema}{Night}
\Delta SolarPanel
\where
slrState=partialEnergyGeneration \lor fullEnergyGeneration\\
slrState'=noEnergyGeneration

\end{schema}

The success and failure conditions for the solar's schemas is summarized in table \ref{tbl:ftypeslr}.
\begin{table*}[h]
\centering
\caption{ Success and error outcomes of the schemas for a solar entity }
\label{tbl:ftypeslr}
\begin{tabular}{| l | l | l |}\hline
\textbf{Schema} & \textbf{Pre-condition for success} &\textbf{Condition for error}\\ \hline  

DayAndCloudy & \makecell{Solar is in no generation state\\$slrState=noGeneration$} & \makecell{Already in partial generation state\\ $slrState=$ \\ $partialEnergyGeneration$}  \\

\hline

DayAndSunny & \makecell{Solar is in partial state \\ 
 $slrState=$\\$partialEnergyGeneration$} &\makecell{Already in full generation state \\
$slrState=$ \\ $fullEnergyGeneration$}
\\ \hline 

Night & \makecell{Solar is in partial or full state
\\ 
$slrState=$\\$PartialEnergyGerention\lor $ \\ $fullEnergyGeneration$
\\}  & \makecell{Already in no generation state \\
$slrState=$\\$noEnergyGeneration$
}  \\

\hline
 
\end{tabular}
\end{table*} 
\subsection{Storage system}

\begin{figure}[h]
\begin{center}
  \includegraphics[width=8.0 cm, height=10.0 cm]{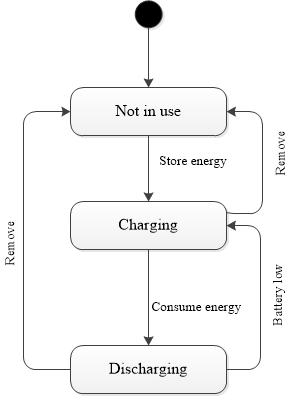}
  \caption{A storage lifetime}\label{fig:storagelifetime}
\end{center}
\end{figure}

A storage device also has different states as well as events that cause the transition of states. As we can see in the figure \ref{fig:storagelifetime} that there are three states named as ``charging, discharging, and not in use'' of a storage object. When a store energy or battery low event occurs, the storage enters in the charging state. In case of energy consumption, the state of storage becomes discharging, In case remove event, the state of storage becomes not in use.

Next, we present formal specification for a storage system.

First, we start by defining a free type \emph{STORAGESTATE} set that comprises of different state of a storage device.
\begin{zed}
[STORAGESTATE]==\{charging,discharging,notInUse\}
\end{zed}
Here, we define a storage system schema. In the schema, we declare a variable ``strState'' of type \emph{STORAGESTATE} that has any single value from the storage state set.
\begin{schema}{StorageDevice}
strState:STORAGESTATE
\end{schema}
As we have defined storage state and system schema. Now, we move towards operational schema.

First, we start by introducing an initialization schema called \emph{InitSorage}. This schema calls the \emph{StorageDevice} schema as a variable. In predicate, it shows that the state of a storage is ``notInUse'' at the initial time.
\begin{schema}{InitStorage}
StorageDevice\\
\where
strState=notInUse

\end{schema}
As we noted before, when a store energy event occurs, a storage goes to the charging state. This process is shown by \emph{StoreEnergy} schema. The changing of a storage is shown by a delta sign. In predicate, it shows that the current state of a storage is ``notInUse'' which is changed to the ``charging'' state.
\begin{schema}{StoreEnergy}
\Delta StorageDevice
\where
strState=notInUse\\
strState'=charging
\end{schema}

We also know that when a consume energy event occurs, a storage device goes to the discharging state. This process is shown by presenting \emph{ConsumeEnergy} schema. In the schema, we can see that the state of a storage changes from charging to the discharging state. 
\begin{schema}{ConsumeEnergy}
\Delta StorageDevice
\where
strState=charging\\
strState'=discharging
\end{schema}

A storage can also enter in the charging state whenever the storage state is low. This process is presented by means of \emph{BatteryLow} schema.
\begin{schema}{BatteryLow}
\Delta StorageDevice
\where
strState=discharging\\
strState'=charging
\end{schema}
When a storage is removed from the system, then it goes to the ``notInUse'' state. This process is shown by introducing another schema called \emph{RemoveStorage}. In the schema, we can see a storage can be in either charging or discharging state. Then it's state changes to the ``notInUse'' state.
\begin{schema}{RemoveStorage}
\Delta StorageDevice
\where
strState=charging\lor discharging\\
strState'=notInUse

\end{schema}

The schemas of the storage devices can be success and may be there some errors at the execution time. So there is also need to define the success and error criteria for schemas. Here, in table \ref{tbl:ftypestr}, we summarized all success and failure conditions of the storage's schemas.

\begin{table*}[h]
\centering
\caption{ Success and error outcomes of the schemas for a wind turbine entity }
\label{tbl:ftypestr}
\begin{tabular}{| c | c | c |}\hline
\textbf{Schema} & \textbf{Pre-condition for success} &\textbf{Condition for error}\\ \hline  

StoreEnergy & \makecell{Storage is in not-in use\\$strState=notInUse$} & \makecell{Already in charging state\\ $strState=charging$}  \\

\hline

ConsumeEnergy & \makecell{Storage is in charging state \\ 
 $strState=charging$} &\makecell{Already in discharging state \\
$strState=discharging$}
\\ \hline 

BatteryLow & \makecell{Storage is in discharging state
\\ 
$strState=discharging$
\\}  & \makecell{Already in charging state \\
$strState=charging$
}  \\

\hline
 
Remove & \makecell{Storage is in charging or dischargig state
\\ 
$strState=discharging \lor charging$
\\}  & \makecell{Already in not-in-use state \\
$strState=notInUse$
}  \\

\hline

\end{tabular}
\end{table*}

\section{Conclusions \& Future Work}

In this paper, we focused on formal specification of smart grid components. We considered four different components of a smart grid system. We identified states and events which cause state transitions of each individual component. Then we modeled each component of the smart grid system by using a formal specification approach. This provides a clear understanding the behavior of each component involved in the system. 
The presented formal framework clearly demonstrates the utility of using a formal specification approach to modeling complex systems including technological systems such as the smart grid.
Here we would like to mention that the presented framework currently considers only four key smart grid components. However, the same can be further expanded for a large number of devices. Additionally, in the future, the work can be expanded to formally model using other approaches such as Petri nets, among others.

\newpage
\ifCLASSOPTIONcaptionsoff
  \newpage
\fi
\bibliography{sample}
\bibliographystyle{IEEEtran}

\end{document}